\documentclass[aps,showpacs,preprintnumbers,amsmath,amssymb,nofootinbib]{revtex4}

\usepackage{graphicx}
\usepackage{color}
      
\def \be  {\begin{equation}}
\def \ee  {\end{equation}}
\def \ee  {\end{equation}}
\def \bea {\begin{eqnarray}}
\def \eea {\end{eqnarray}}

\def \Tr  {\bf{Tr}}

\newcommand{\ba} {\begin{eqnarray}}
\newcommand{\ea} {\end{eqnarray}}

\begin{document}

\preprint{ECTP-2012-06}

\title{Particle Production at RHIC and LHC Energies}

\author{A.~Tawfik}
\email{a.tawfik@eng.mti.edu.eg}
\email{atawfik@cern.ch}
\affiliation{Egyptian Center for Theoretical Physics (ECTP), MTI University, Cairo, Egypt}

\author{E. Gamal}
\affiliation{Egyptian Center for Theoretical Physics (ECTP), MTI University, Cairo, Egypt}

\author{A. G. Shalaby}
\affiliation{Physics Department, Faculty of Science, Benha University, Benha 13518, Egypt}

\date{\today}

\begin{abstract}
The production of different particle species is recently measured in $Pb-Pb$ collisions by the ALICE experiment at $\sqrt{s}=7~$TeV. This motivates the use of various bosons and baryons measured at lower center-of-mass energies in comparing their ratios to the hadron resonance (HRG) gas model and PYTHIA event generator. It is found that the particle-to-antiparticle ratios are perfectly reproduce by means of HRG and PYTHIA at RHIC and LHC energies. The kaon-to-pion and proton-to-pion ratios are entirely overestimated by the HRG model. The PYTHIA event generator obviously underestimates the kaon-to-pion ratio and simultaneously reproduces the proton-to-pion ratio, almost perfectly, especially at LHC energy.
While matter-to-antimatter and non-strange abundances are partly in line with predictions from the HRG model, it is found in the ALICE experiment that the measured baryon ratios are suppressed by a factor of $\sim1.5$. The strange abundances are overestimated in the HRG model. 

\end{abstract}


\maketitle


\section{Introduction}
\label{sec:intr}

Understanding the dynamical properties of hot and dense hadronic matter is one of the main motivations of heavy-ion experiments, which in turn offer unique possibilities to study hadronic matter under extreme conditions \cite{pp1,pp2,pp3a,pp3b} and to compare with the lattice QCD simulations \cite{pp4}. The Relativistic Heavy-Ion Collider (RHIC) has shown that the bulk matter created in such collisions can be quantitatively described by hydrodynamic models \cite{pp3b}. The created hot and dense partonic matter, the quark-gluon plasma (QGP) or colored quarks and gluons, where  quarks and gluons can move freely over large volumes comparing to the typical size of a hadron, rapidly expands and cools down. Over this path, it likely undergoes phase transition(s) back to the hadronic matter. Different thermal models can very well reproduce the produced particle abundances, which are governed in chemical equilibrium by two parameters, the chemical freeze-out temperature $T_{ch}$ and the baryochemical potential $\mu_b$, where the latter reflects the net baryon content of the system,\, directly, and the center-of-mass energy, indirectly.

That the particle abundances at Alternating Gradient Synchrotron (AGS), Super Proton Synchrotron (SPS) and RHIC energies are consistent with equilibrium populations \cite{therm1} makes it possible to extract both freeze-out parameters over a wide range of center-of-mass energies $\sqrt{s_{NN}}$ from fits of measured particle ratios with thermal models, which obviously do not count for hadrons interactions in the final state. Nevertheless, the formation of resonances can only be materialized through strong interactions since the resonances (fireballs) are composed of further resonances (fireballs), which in  turn consist of resonances (fireballs) and so on \cite{hged}. Taking into consideration all kinds of resonance interactions by means of the S-matrix, which describes the scattering processes in the thermodynamical system, reduces the resulting virial term, so that the partition function turns to be reduced to the non-relativistic limit, especially at narrow width and/or low temperature $T$ \cite{Tawfik:2004sw}.  The resonance contributions to the partition function are the same as that of collisionless particles with some effective mass. All possible interactions modifying the relative abundances are found negligible in the hadronic phase \cite{intr1,Tawfik:2004sw}. It is assumed that the hadron resonances with masses $<2.8~$ GeV avoid the singularities expected at Hagedorn temperature \cite{Karsch:2003vd,Karsch:2003zq,Redlich:2004gp,Taw3}.

The earliest idea about enhancement of strangeness abundances as a key signature of QGP formation in heavy-ion collisions has been proposed, three decades ago \cite{str1}. During the hadronization process, $T$ drops from about two hundred to few tens MeV, the dynamical mass of strange quark drops, as well. As a consequence, strangeness in QGP would equilibrate on small time scales relative to those in a hadronic matter \cite{str2}. At SPS energy, there were strong enhancements observed in AA- and almost neglecting ones in pA-collisions. This has been cited as an experimental evidence for QGP formation \cite{sps}. The enhancements at  AGS energy are considered a typical rescattering of produced hadrons.  Assuming that thermally equilibrated QGP hadronizes into a maximum entropy state, a test for strange quark saturation in the early stages is provided by comparing final state hadron yields to thermal model predictions \cite{str3,star2012}.

Studying the ratios of particle yields helps in determining the freeze-out parameters and in eliminating the volume fluctuations. Furthermore, the dependence of the freeze-out surface on the initial conditions can be neglected. Using statistical model with corresponding choices of the thermal  model parameters with $\sqrt{s_{NN}}$ made according to the systematics extracted from heavy-ion collisions at lower energies, some predictions of particle abundances at LHC energy are reported \cite{henrg1a,henrg1b}. Using the hadron resonance gas (HRG) model, $\bar{p}/p$ and $K^-/K^+$ are given in dependence on $\sqrt{s_{NN}}$ \cite{Tawfik:2010pt,Tawfik:2010kz}. The present work is motivated by recent measurements of pion, kaon and proton ratios in central Pb-Pb collisions at $\sqrt{s_{NN}}=2.76~$ TeV by the ALICE experiment. 

The present paper is organized as follows. Section \ref{sec:mdl} elaborates details on HRG model and the event generator PYTHIA. The results and discussions are outlined in section \ref{sec:res}. The conclusions are discussed in section \ref{sec:conc}. 

\section{The Model}
\label{sec:mdl}

\subsection{Hadron Resonance Gas Model}

Treating the hadron resonances as a free gas is supposed to construct the thermodynamic pressure in the hadronic phase i.e., $<T_c$ \cite{Karsch:2003vd,Karsch:2003zq,Redlich:2004gp,Taw3}. This is valid for collisionless and interacting hadron resonances. It has been shown that the thermodynamics of strongly interacting system can also be approximated to an ideal gas composed of hadron resonances with masses $\le 2~$GeV ~\cite{Tawfik:2004sw,Vunog}. Therefore, the confined phase of QCD, the hadronic phase, would be modelled as a non-interacting gas of hadron resonances. The grand canonical partition function can be given as: 
\bea
Z(T,V) &=&\Tr\left[ \exp^{-\mathbf{H}/T}\right],
\eea

where $\mathbf{H}$ is the Hamiltonian of the system, which is given by the summation over the  kinetic energies of relativistic Fermi and Bose particles. The main motivation of using this Hamiltonian is that it contains all relevant degrees of freedom of confined, strongly interacting matter. It includes implicitly the interactions that result in the formation of the resonances. In addition, it has been shown that this model gives a quite satisfactory description of the particle production in the heavy-ion collisions. With the above assumptions, the dynamics of the partition function can be calculated exactly and be expressed as summation over 
{\it single-particle partition} functions $Z_i^1$ of all hadrons and their resonances.
\bea\label{eq:lnz1}
\ln Z(T, \mu ,V)&=&\sum_i \ln Z^1_i(T,V)=\sum_i\pm \frac{V g_i}{2\pi^2}\int_0^{\infty} k^2 dk \ln\left\{1\pm \exp[(\mu_i -\varepsilon_i)/T]\right\},
\eea
where $\varepsilon_i(k)=(k^2+ m_i^2)^{1/2}$ is the $i-$th particle dispersion relation, $g_i$ is
spin-isospin degeneracy factor and $\pm$ stands for bosons and fermions, respectively.

Before the discovery of QCD, it was speculated about a possible phase transition of a massless pion gas to a new phase of matter. Based on statistical models like Hagedorn \cite{hgdrn1} and Bootstrap \cite{boots1}, the thermodynamics of such an ideal pion gas is studied, extensively. After the QCD, the new phase of matter is known as QGP. The physical picture was that at $T_c$ the additional degrees of freedom carried by QGP are to be released resulting in an increase in the thermodynamic quantities. The success of HRG in reproducing lattice QCD results at various quark flavours and masses (below $T_c$) changed this physical picture, drastically. Instead of releasing additional degrees of freedom at $T>T_c$, it is found that the interacting system reduces its effective degrees of freedom at $T<T_c$. In other words, the hadron gas has much more degrees of freedom than QGP.

At finite temperature $T$ and baryo-chemical potential $\mu_i $, the pressure of $i$-th hadron or resonance reads 
\begin{eqnarray}
\label{eq:lnz1} 
p(T,\mu_i ) &=& \pm \sum_i^{N} \frac{g_i}{2\pi^2}T \int_{0}^{\infty}
           k^2 dk  \ln\left\{1 \pm \exp[(\mu_i -\varepsilon_i)/T]\right\},
\end{eqnarray}
where $N$ is the total number of hadron resonances of interest. 
As no phase transition is conjectured in HRG, summing over all hadron resonances results in the final thermodynamic pressure in the hadronic phase. 
Switching between hadron and quark chemistry is given by the correspondence between  the {\it hadronic} chemical potentials and that of the quark constituents, for example, 
$\mu_i =3\, n_b\, \mu_q + n_s\, \mu_S$, where $n_b$($n_s$) being baryon (strange) quantum number. The chemical potential assigned to the {\it degenerate} light quarks is $\mu_q=(\mu_u+\mu_d)/2$ and the one assigned to strange quark reads $\mu_S=\mu_q-\mu_s$. 

The strangeness chemical potential $\mu_S$ is treated as a dependent parameter. Basically, it is 
calculated as a function of $T$ and $\mu_b$, based on the fact that the overall strange quantum number has to remain conserved in heavy-ion collisions~\cite{Tawfik:2004sw}.  Therefore and in order to assure vanishing strange charge, $\mu_S(\mu_b,T)$ has to be calculated, whenever  baryochemical potential $\mu_b$ and/or temperature $T$ are changed.

When ignoring all decay channels, the particle number density is given by the derivative of the partition function, Eq.~\ref{eq:lnz1},  with respect to the chemical potential of interest. 
\ba \label{eq:n1} 
\langle n\rangle &=& \sum_i \frac{g_i}{2\pi^2} \int dk k^2
\frac{e^{(\mu_i-\varepsilon_i)/T}}{1\pm e^{(\mu_i-\varepsilon_i)/T}}.
\ea
In the present work,  $T$ and $\mu_b$, at which the chemical freeze-out takes place, are characterized by constant  $s/T^3$, where $s$ is the entropy density. Details about this ratio and its physical meaning are given in \cite{Taw3}.

\subsection{PYTHIA}

Although the experimental data shown in the present work are taken from heavy-ion collisions, the comparison with PYTHIA, which is designed to generate multiparticle production in collisions between elementary particles, $e^+ e^-$, $pp$ and $ep$, is possible at very high energies.
The bulk of PYTHIA multiplicities is formed in jets, i.e. in collimated bunches of hadrons or resonances decaying into further hadrons produced by the hadronization of partons \cite{pythia1}. 
The relative proportion of strange particles is as expected smaller in comparison with nonstrange hadrons. PYTHIA is capable of simulating for different processes including hard and soft interactions, parton distributions, initial/final-state parton showers, multiple interactions, fragmentation and decay.

The measured particle production is conjectured as an indicator for the formation of QGP, especially in heavy-ion collisions. In $pp$ collisions, the spatial and time evolution of the system is too short to assure initial conditions required to drive hadronic matter into partonic QGP. The HRG model has been successfully used to describe particle abundances and their fluctuations in heavy-ion collisions. The thermodynamics of lattice QCD is also reproduced by means of HRG at temperatures below the critical one \cite{Karsch:2003vd,Karsch:2003zq,Redlich:2004gp,Taw3}. 

In Refs. \cite{Tawfik:2010pt,Tawfik:2010kz}, we have noticed that the collective flow of strongly interacting matter in heavy-ion collisions makes HRG model underestimating the particle yields ratios measured in $pp$ collisions, especially at low energies. Nevertheless, it was found that the differences between particle yields ratios of $pp$ and $AA$ collisions almost disappear, at LHC energies. In light of this, the comparison with PYTHIA remains an enlightening feature. Although, it gives comparable high-energy results as the ones from heavy-ion collisions, its initial conditions would be reflected in the collective properties in the final state. This would include - among other - the issue of strangeness suppression, as the mass of strange quark is heavier than that of up and down quarks. Therefore, the production of strange hadrons is generally suppressed relative to hadrons containing only up and down quarks.

\section{Results and Discussions}
\label{sec:res}

\begin{figure}[htb!]
\includegraphics[width=7.cm,angle=-90]{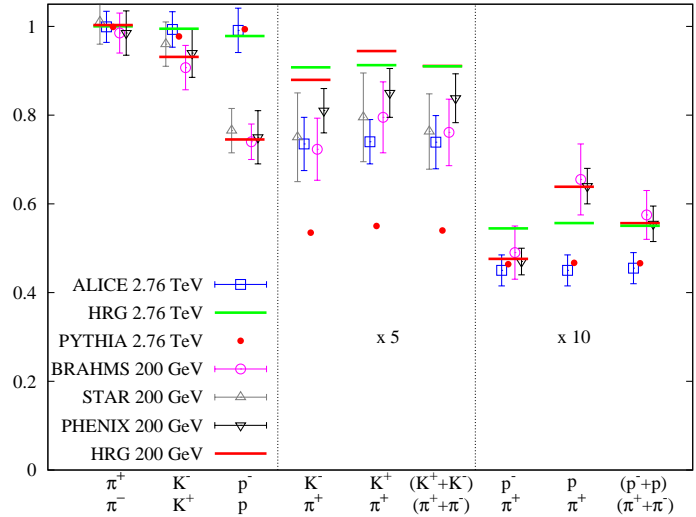}
\caption{Measured particle ratios (symbols) are compared to the predictions from HRG model (horizontal lines) and PYTHIA even generator (solid circles).}
\label{fig:1a}
\end{figure}

In Fig. \ref{fig:1a}, the ratios of pion, kaon and protons at RHIC and LHC energies (open symbols) are compared with the HRG model (horizonal lines) and the PYTHIA event generator (solid circle). The HRG results are calculated at $200~$GeV corresponding to the RHIC energy and $2.76~$TeV corresponding to the LHC energy. The PYTHIA simulations are performed for $pp$ collisions at $2.76~$TeV. First, we study the ratios of particle-to-antiparticle given in the left panel. It is obvious, that the experimental results are well reproduced by means of both HRG and PYTHIA. Furthermore, the energy-dependence is very well reflected. The right panel shows the different proton-to-pion ratios.  We notice that PYTHIA results describe very well the ALICE experimental results taken at $2.76~$TeV. It is apparent that the HRG results on $\bar{p}/\pi$ nearly reproduce the experimental data, while the ratios of $p/\pi$ and $(\bar{p}+p)/(\pi^++\pi^-)$ obviously underestimate the corresponding ALICE results. The RHIC results on the three ratios are very well reproduced by means of the HRG model. It is interesting to notice that the middle panel gives amazingly different results. Both experimental data sets are not clearly distinguishable, although the huge jump in center-of-mass energy $\sqrt{s_{NN}}$ from RHIC ($200~$GeV) to LHC ($2760~$GeV). We notice that the PYTHIA event generator entirely underestimates the experimental results, while the HRG model overestimates them.

Fig. \ref{fig:1a} shows a collective comparison. Another comparative study is summarized in Tab. \ref{tab1}, where the HRG results on particle-antiparticle ratios are compared with the corresponding results reported in Ref. \cite{henrg1a}. The mixed particle ratios are listed in Tab. \ref{tab2}. It compares the HRG results with previous studies \cite{henrg1a,henrg1b}.
\begin{table}[h]
\begin{tabular}{|l||c|c|c|c|c|c|}\hline
  & $\frac{K^-}{K^+}$ & $\frac{\pi^+}{\pi^-}$ & $\frac{\bar{p}}{p}$ & $\frac{\bar{\Lambda}}{\Lambda}$ & $\frac{\bar{\Xi}^+ }{\Xi^-}$ & $\frac{\Omega^+}{\Omega^-}$  \\ \hline \hline
 HRG &$0.994$ & $0.9997$& $0.978$&$0.983$&& \\ \hline 
Ref. \cite{henrg1a} &$0.9998$&$0.9998$&$0.989$&$0.992$&$0.994$&$0.996$\\ \hline 
\end{tabular} 
\caption{Ratios of matter-antimatter compared with the results reported in Ref. \cite{henrg1a}. \label{tab1}}
\end{table}

\begin{table}[h]
\begin{tabular}{|l||c|c|c|c|c|c|c|c|c|}\hline
 & $\frac{K^+}{\pi^+}$ & $\frac{K^-}{\pi-}$ & $\frac{p}{\pi^-}$ & $\frac{p}{\pi^+}$ &$\frac{\Lambda}{p}$ & $\frac{\Lambda}{\pi^-}$ &$\frac{\Xi^-}{\Lambda}$ & $\frac{\Xi^-}{\pi^-}$ & $\frac{\Omega^-}{\Xi^-}$ \\ \hline \hline
 HRG &$0.182$&$0.181$&&$0.056$&$0.437$&$0.024$&&& \\ \hline 
Ref. \cite{henrg1a} &$0.180$&$0.179$&$0.091$&&$0.473$&&$0.160$&&$0.186$ \\ \hline 
Ref.  \cite{henrg1b} &$0.164$&$0.163$&& $0.072$ &&$0.042$&&$0.0054$&$0.00093$ \\
\hline
\end{tabular} 
\caption{Ratios of mixed particle species compared with the results reported in Ref. \cite{henrg1a} and \cite{henrg1b} and compared with experimental measurements at RHIC and LHC. \label{tab2} }
\end{table}

\begin{figure}[htb!]
\includegraphics[width=7.cm,angle=-90]{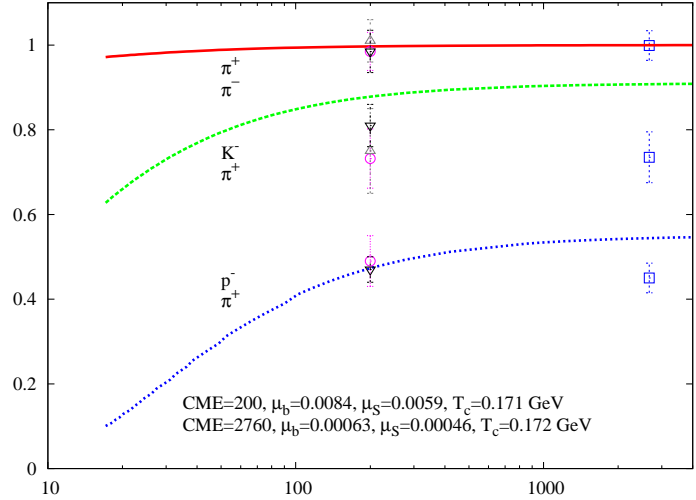}
\caption{Ratios of $\pi^+/\pi^-$, $K^-/\pi^+$ and $\bar{p}/\pi^+$ are given in dependence on $\sqrt{s_{NN}}$ and compared with experimental measurements at RHIC and LHC. \label{fig:2a}}
\end{figure}

\begin{figure}[htb!]
\includegraphics[width=7.cm,angle=-90]{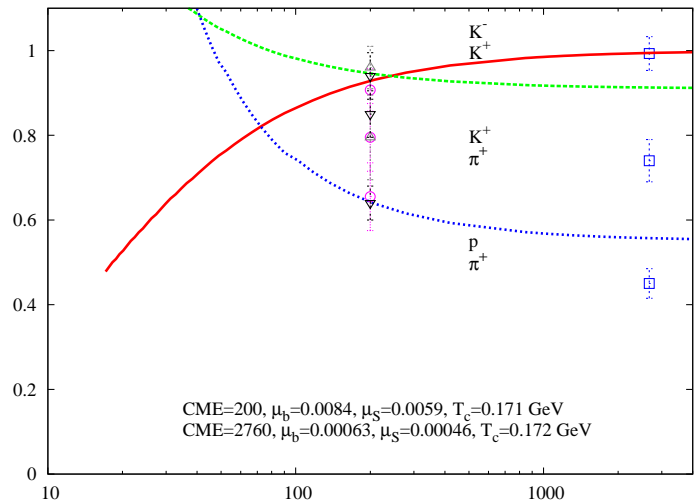}
\caption{Ratios of $K^-/K^+$, $K^+/\pi^-$,  and $p/\pi^+$ are given in dependence on $\sqrt{s_{NN}}$ and compared with experimental measurements at RHIC and LHC. \label{fig:3a} }
\end{figure}

\begin{figure}[htb!]
\includegraphics[width=7.cm,angle=-90]{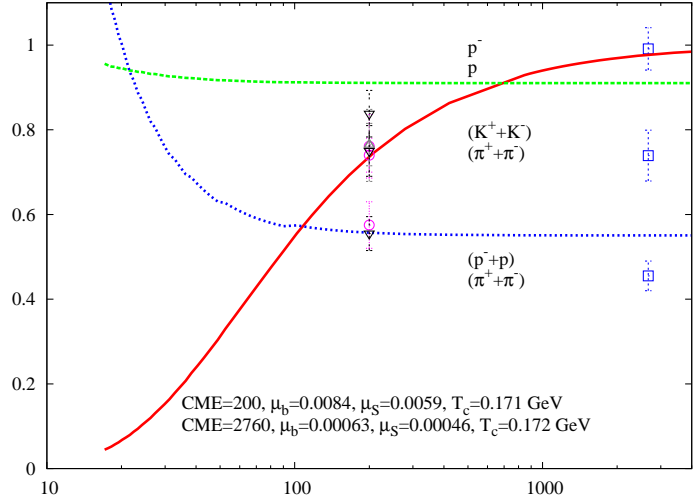}
\caption{Ratios of $\bar{p}/p$, $(K^++K^-)/(\pi^++\pi^-)$,  and $(\bar{p}+p)/(\pi^++\pi^-)$ are given in dependence on $\sqrt{s_{NN}}$ and compared with experimental measurements at RHIC and LHC. \label{fig:4a} }
\end{figure}

A quantitative comparison is illustrated in Fig. \ref{fig:2a}, \ref{fig:3a} and \ref{fig:4a}. In Fig. \ref{fig:2a}, $\pi^+/\pi^-$, $K^-/\pi^+$ and $\bar{p}/\pi^+$ ratios in dependence on $\sqrt{s_{NN}}$. The solid curve represents the HRG calculations for $\pi^+/\pi^-$. As shown in Fig. \ref{fig:1a}, the agreement with both experimental data sets (RHIC and LHC) is excellent. This has been introduce in Ref. \cite{Tawfik:2010pt,Tawfik:2010kz}. The dashed curve represents the energy evolution of $K^-/\pi^+$ ratio. Here, the experimental data lay below the experimental ones. The third ratio is given by the dotted curve. We find that the HRG model reproduces the RHIC results for $\bar{p}/\pi^+$, for which the ALICE results are overestimated.

In Fig. \ref{fig:3a}, the strange bosonic ratio of $K^-/K^+$, strange to non-strange bosonic ratio of  $K^+/\pi^+$,  and baryon to non-strange boson $p/\pi^+$ ratios are given in dependence on $\sqrt{s_{NN}}$. The solid curve represents the HRG calculations for anti-kaon to kaon ratio $K^-/K^+$. As shown in Fig. \ref{fig:1a}, the agreement with both experimental data sets (RHIC and LHC) is excellent \cite{Tawfik:2010pt,Tawfik:2010kz}. The particle ratio of $K^+/\pi^+$ is represented by the dashed curve. It slightly decreases with increasing $\sqrt{s_{NN}}$. Once again, the HRG model obviously overestimates the experimental results, entirely. The dotted curve represents the HRG results on $p/\pi^+$. We find that the RHIC results are well described by the HRG model, while the ALICE data are obviously overestimated.

Finally, the three remaining ratios $\bar{p}/p$, $(K^++K^-)/(\pi^++\pi^-)$ and $(\bar{p}+p)/(\pi^++\pi^-)$ are given in Fig. \ref{fig:4a} as function of $\sqrt{s_{NN}}$.  
Once again, the agreement between the experimental results (RHIC and LHC) and the HRG model on $\bar{p}/p$ is excellent, Fig. \ref{fig:1a} and Ref. \cite{Tawfik:2010pt,Tawfik:2010kz}. The kaon-to-pion ratio is overestimated by the HRG model, while the proton-to-pion ratio is well reproduced at the RHIC energy but once again overestimated at the LHC energy. The kaon-to-pion ratios are entirely overestimated. Along the entire energy range, the HRG model gives almost $30\%$ higher values than the measured ratios.

\section{Conclusions}
 \label{sec:conc}
 
Based on the study given in section \ref{sec:res}, two remarks are now in order. The first one is that the energy scan of PYTHIA is completely excluded, because of its basic assumptions. The PYTHIA simulations are performed for $pp$ collisions. Therefore, the comparison with the heavy-ion collisions would not helpful, especially at low energies. The second remark is that the HRG model counts for all baryonic and bosonic resonances with masses $<2.8~$GeV, assures conserved strange degrees of freedom, excludes all decay channels and finally characterizes the chemical freeze-out by constant $s/T^3$ over the whole energy range.   

The number densities of different particle species are calculated in framework of the HRG model and compared with the experimental results at RHIC and LHC energies. Fig. \ref{fig:1a} summarizes the results, collectively. We notice that the particle-to-antiparticle ratios are best described by the HRG model and PYTHIA event generator at both RHIC and LHC energies.  The results on mixed particle ratios are different. Obviously, the strange particles are not reproduced by the HRG model. At RHIC and LHC energies all ratios including kaon-mesons are overestimated. The PYTHIA event generator apparently underestimates all these ratios, especially at $2.76~$TeV. 
The proton-to-pion ratios are partially described by the HRG model, at RHIC energy, while the ALICE results are apparently overestimated. Obviously, PYTHIA reproduces very well all these ratios, especially at $2.76~$TeV. 

So far, we conclude that the particle-antiparticle ratios are very well reproduced by means of HRG and PYTHIA. These two models are partially able to reproduce the mixed particles ratios. This appears very clearly when comparing their results with the experimental kaon-to-pion and proton-to-pion ratios. The strange ratios quark flavours seem to play an essential role in explaining the discrepancy with the kaon-to-pion ratios. On one hand, the HRG model, which excludes all decay channels, seems to overestimate the experimental results. On the other hand, the PYTHIA event generator, which is basically originated from $pp$ interactions, whose spatial and temporal evolutions are likely short to assure initial conditions driving the hadronic matter to the partonic QGP, seems to underestimate the experimental results. 
The threshold for strange quark production in QGP is much smaller than in hadronic matter. This effect would be further enhanced in baryons with multiple strange quarks. While QGP is more likely to be found in heavy-ion collisions, the strangeness enhancement in high energetic $pp$ collisions would be a sign of a collective effect \cite{se1}.

So far, there are three main features of ALICE measurements, Fig. \ref{fig:1a}. The first one is that the particle-to-antiparticle we very well reproduced by means of the HRG model and the PYTHIA event generator at RHIC and LHC energies. The second one is that the kaon-to-pion and proton-to-pion ratios are entirely overestimated by the HRG model. The third one is that the PYTHIA event generator obviously underestimates the first ratio and simultaneously reproduces the second ratio, almost perfectly.

The huge suppression of strange quark production in PYTHIA would be originated in the absence of correct inclusion of strange production in current tunes of PYTHIA. It has been concluded that there is a large increase in the measured production cross section of strange particles as the energy increases from $0.9$ to $7$ TeV \cite{se2}. Also, it is found that he difference between predictions of strange particles production and measurements gets bigger as the particle mass and strangeness number increase.




\begin{thebibliography}{99}

\bibitem{pp1} N. Cabibbo and G. Parisi, Phys.Lett. B59, 67 (1975).

\bibitem{pp2} E. V. Shuryak, Phys.Lett. B78, 150 (1978).

\bibitem{pp3a} L. D. McLerran and B. Svetitsky, Phys.Lett. B {\bf 98}, 195 (1981).

\bibitem{pp3b} M. Gyulassy and L. D. McLerran,  Nucl.Phys. A {\bf 750}, 30-63 (2005).

\bibitem{pp4} Z. Fodor, Nucl. Phys. A {\bf 715}, 319-328 (2003).

\bibitem{therm1} A. Andronic, P. Braun-Munzinger, K. Redlich, and J. Stachel, J. Phys. G {\bf G38}, 124081 (2011).

\bibitem{hged} R. Hagedorn, Nuovo Cim. Suppl. {\bf 3}, 147 (1965).

\bibitem{Tawfik:2004sw} A. Tawfik, Phys.Rev. D {\bf 71}, 054502 (2005). 

\bibitem{intr1} R. Rapp and E. V. Shuryak, Phys. Rev. Lett. {\bf 86}, 2980 (2001).

\bibitem{Karsch:2003vd}F.~Karsch, K.~Redlich and A.~Tawfik, 
            Eur.~Phys.~J.~C {\bf 29},~549~(2003).

\bibitem{Karsch:2003zq}F.~Karsch, K.~Redlich and A.~Tawfik,
           Phys.~Lett.~B {\bf 571},~67~(2003).

\bibitem{Redlich:2004gp}K.~Redlich, F.~Karsch and  A.~Tawfik,
           J.~Phys.,~G {\bf 30},~S1271~(2004). 

\bibitem{Taw3} A.~Tawfik, J.~Phys.~G {\bf 31}~S1105,~(2005).

\bibitem{str1}  J. Rafelski and B. M\"uller, Phys. Rev. Lett. {\bf 48}, 1066 (1982).

\bibitem{str2}  P. Koch, B. M\"uller, and J. Rafelski, Phys. Rep. {\bf 142},  167 (1986). 

\bibitem{sps} U. Heinz, Nucl. Phys. A {\bf 685},  414-431 (2001). 

\bibitem{str3} S. Hamieh, K. Redlich and A. Tounsi, Phys. Lett. B {\bf 486}, 61 (2000).

\bibitem{star2012} G. Agakishiev {\it et al.} [STAR Collaboration], Phys. Rev. Lett. {\bf 108}, 072301  (2012).

\bibitem{henrg1a} J. Cleymans, I. Kraus, H. Oeschler, K. Redlich and S. Wheaton, Phys.Rev. C {\bf 74}, 034903 (2006).

\bibitem{henrg1b} A. Andronic, P. Braun-Munzinger and  J. Stachel, Phys. Lett. B {\bf 673}, 142-145 (2009). 

\bibitem{Tawfik:2010pt} A. Tawfik, Nucl.Phys. A {\bf 859}, 63-72 (2011).

\bibitem{Tawfik:2010kz} A. Tawfik, Int. J. Theor. Phys. {\bf 51}, 1396-1407 (2012).

\bibitem{Vunog} R.~Venugopalan, M.~Prakash, Nucl.~Phys.~A {\bf 546},~718~(1992). 

\bibitem{hgdrn1} R. Hagedorn, Nuovo Cim. Suppl. {\bf 6}, 311-354 (1968); Nuovo Cim. A {\bf 56}, 1027-1057 (1968). 

\bibitem{boots1} R.J. Eden, P.V. Landshoff, D.I. Olive and J.C. Polkinghorne, {\it The Analytic S-Matrix}, Cambridge University Press, 1966; 
J. Letessier‏, J. Rafelski‏, {\it Hadrons and quark-gluon plasma}, Cambridge University Press, 2002. 

\bibitem{pythia1} T. Sj\"ostrand, S. Mrenna and P. Skands, JHEP, {\bf 05}, 026 (2006).

\bibitem{se1} V. Khachatryan, {\it et al.} [CMS Collaboration], JHEP {\bf 1105}, 064 (2011).

\bibitem{se2} S. Lobanov, A. Maevskiy and L. Smirnova, PoS (IHEP-LHC-2011), 008 (2011). 

\end{thebibliography}
\end{document}